# Approaching the parameter estimation quality of maximum likelihood via generalized moments

*Fyodor V. Tkachov*
*Institute for Nuclear Research*
*of Russian Academy of Sciences*
*Moscow 117312 Russia*

A simple criterion is presented for a practical construction of generalized moments that allow one to approach the theoretical Rao-Cramer limit for parameter estimation while avoiding the complexity of the maximum likelihood method in the cases of complicated probability distributions and/or very large event samples.

INTRODUCTION. The purpose of this note is to describe a result that was discovered in a rather special context of the theory of so-called jet finding algorithms [1] but seems to be basic enough to belong to the core statistical wisdom of parameter estimation.

Namely, I would like to present a simple formula (Eq.(20)) that connects the method of generalized moments with the maximum likelihood method by explicitly describing deviations from the Rao-Cramer limit on precision of parameter estimation with a given event sample; see e.g. [2], [3].

The formula leads to practical prescriptions (the method of quasi-optimal moments[a]; see after Eq.(24)) that offer a practical alternative to the maximum likelihood method in precision measurement problems when the use of the maximum likelihood method is impractical due to complexity of theoretical expressions for the probability distribution or a large size of the sample of events.

Although closely related to the well-known results and mathematical techniques, the prescription is new to the extent that I've seen no trace in the literature of its being known to physicists despite its immediate relevance to precision measurements.

THE PROBLEM. One deals with a random variable $\mathbf{P}$ whose instances (specific values) are called events. Their probability density is denoted as $\pi(\mathbf{P})$. It is assumed to depend on a parameter $M$ which has to be estimated from an experimental sample of events $\{\mathbf{P}_i\}_i$.

The standard method of generalized moments consists in choosing a function $f(\mathbf{P})$ defined on events (the generalized moment), and then finding $M$ by fitting its theoretical average value,

$$\langle f \rangle = \int d\mathbf{P}\, \pi(\mathbf{P}) f(\mathbf{P}) , \qquad (1)$$

against the corresponding experimental value:

$$\langle f \rangle_{\exp} = \frac{1}{N} \sum_i f(\mathbf{P}_i) . \qquad (2)$$

The problem is to find $f$ which would allow one to extract $M$ with the highest precision from the event sample.

OPTIMAL MOMENTS. In the context of precision measurements one can assume the magnitude of errors to be small. Then fluctuations in the values of $M$ are related to fluctuations in the values of $\langle f \rangle$ as

$$\delta M = \left(\frac{\partial \langle f \rangle}{\partial M}\right)^{-1} \delta \langle f \rangle . \qquad (3)$$

The derivative is applied only to the probability distribution:

$$\frac{\partial \langle f \rangle}{\partial M} = \int d\mathbf{P}\, f(\mathbf{P}) \frac{\partial \pi(\mathbf{P})}{\partial M} . \qquad (4)$$

This is because $M$ is unknown, so even though the solution, $f_{\text{opt}}$, will depend on $M$, any such dependence is coincidental and therefore "frozen" in this calculation.

For small fluctuations $\delta\langle f \rangle = N^{-1/2} \sqrt{\operatorname{Var}\langle f \rangle}$, where

$$\operatorname{Var}\langle f \rangle = \int d\mathbf{P}\, \pi(\mathbf{P}) (f(\mathbf{P}) - \langle f \rangle)^2 \equiv \langle f^2 \rangle - \langle f \rangle^2 . \qquad (5)$$

In terms of variances, Eq.(3) becomes:

$$\operatorname{Var} M[f] = \left(\frac{\partial \langle f \rangle}{\partial M}\right)^{-2} \operatorname{Var} f . \qquad (6)$$

The problem is to minimized this by a suitable choice of $f$.

A necessary condition for a minimum can be written in terms of functional derivatives:[b]

$$\frac{\delta}{\delta f(\mathbf{P})} \operatorname{Var} M[f] = 0 . \qquad (7)$$

Substitute Eq.(6) into (7) and use the following relations:

---

[a] In the quantum-theoretic context of [1] generalized moments are naturally interpreted as quantum observables, so the method was called the method of quasi-optimal observables.

[b] An interesting mathematical exercise of casting the following reasoning (the functional derivatives, etc.) into a rigorous form is left to interested mathematical parties. A premature emphasis on rigor would have obscured the simple analogy with the study of minima of ordinary functions via the usual Taylor expansion.

For practical purposes it is sufficient to remember that the range of validity of the prescriptions we obtain is practically the same as for the maximum likelihood method. Note that the derivation in terms of functional derivatives can be related to the proofs of the Rao-Cramer inequality in terms of Hilbert statistics, etc.; cf. e.g. [4].

$$\frac{\delta}{\delta f(\mathbf{P})}\langle f \rangle = \pi(\mathbf{P}), \qquad \frac{\delta}{\delta f(\mathbf{P})}\langle f^2 \rangle = 2 f(\mathbf{P}) \pi(\mathbf{P}),$$

$$\frac{\delta}{\delta f(\mathbf{P})}\frac{\partial \langle f \rangle}{\partial M} = \frac{\partial \pi(\mathbf{P})}{\partial M}. \tag{8}$$

After some simple algebra one obtains:

$$f(\mathbf{P}) = \langle f \rangle + \text{const}\, \frac{\partial \ln \pi(\mathbf{P})}{\partial M}, \tag{9}$$

where the constant is independent of $\mathbf{P}$. The constant plays no role since $f$ is defined by this reasoning only up to a constant factor. Noticing that

$$\int d\mathbf{P}\, \pi(\mathbf{P}) \frac{\partial \ln \pi(\mathbf{P})}{\partial M} = \frac{\partial}{\partial M} \int d\mathbf{P}\, \pi(\mathbf{P}) \equiv \frac{\partial}{\partial M} 1 = 0, \tag{10}$$

we arrive at the following general family of solutions:

$$f(\mathbf{P}) = C_1 \frac{\partial \ln \pi(\mathbf{P})}{\partial M} + C_2, \tag{11}$$

where $C_i$ are independent of $\mathbf{P}$ but may depend on $M$.

For convenience of formal investigation we will usually deal with the following member of the family (11):

$$f_{\text{opt}}(\mathbf{P}) = \frac{\partial \ln \pi(\mathbf{P})}{\partial M}. \tag{12}$$

Then Eq. (10) is essentially the same as

$$\langle f_{\text{opt}} \rangle = 0. \tag{13}$$

A SIMPLE EXAMPLE. Consider the familiar Breit-Wigner shape. Let $\mathbf{P}$ be random real numbers distributed according to

$$\pi(\mathbf{P}) \propto \frac{1}{(M-\mathbf{P})^2 + \Gamma^2} \tag{14}$$

in some fixed interval around $\mathbf{P} = M$. Suppose $M$ is unknown. Then the optimal moment is

$$f_{M,\text{opt}}(\mathbf{P}) = \frac{\partial}{\partial M} \ln \pi(\mathbf{P}) = -\frac{2(M-\mathbf{P})}{(M-\mathbf{P})^2 + \Gamma^2}. \tag{15}$$

(Remember that $\mathbf{P}$-independent additive and multiplicative constants can be dropped in such expressions; see Eq. (11).)

It is interesting to observe how $f_{M,\text{opt}}$ emphasizes contributions of the slopes of the bump — exactly where the magnitude of $\pi(\mathbf{P})$ is most sensitive to variations of $M$ — and taking contributions from the two slopes with a different sign maximizes the signal. At the same time the expression (15) suppresses contributions from the middle part of the bump (14) that generates mostly noise as far as $M$ is concerned.

CONNECTION WITH MAXIMUM LIKELIHOOD. Eq. (12) can be regarded as a translation of the method of maximum likelihood (which is known to yield the theoretically best estimate for $M$; cf. the Rao-Cramer inequality [2], [3]) into the language of generalized moments.[c] Indeed, the maximum likelihood method prescribes to estimate $M$ by the value which maximizes the likelihood function,

$$\sum_i \ln \pi(\mathbf{P}_i), \tag{16}$$

where summation runs over all events from the sample. The necessary condition for the maximum of (16) is

$$\frac{\partial}{\partial M}\left[\sum_i \ln \pi(\mathbf{P}_i)\right] = \sum_i \frac{\partial \ln \pi(\mathbf{P}_i)}{\partial M} \propto \langle f_{\text{opt}} \rangle_{\text{exp}} = 0. \tag{17}$$

This agrees with (12) thanks to (13).

DEVIATIONS FROM $f_{\text{opt}}$. Next we are going to consider how small deviations from $f_{\text{opt}}$ affect the precision of extracted $M$. Consider (6) as a functional of $f$, $\text{Var}\,M[f]$. Assume $\varphi$ is a function of events such that $\langle \varphi^2 \rangle < \infty$. We are going to evaluate the functional Taylor expansion of $\text{Var}\,M[f_{\text{opt}} + \varphi]$ with respect to $\varphi$ through quadratic terms:

$$\text{Var}\,M[f_{\text{opt}} + \varphi] = \text{Var}\,M[f_{\text{opt}}]$$
$$+ \frac{1}{2}\int \left[\frac{\delta^2 \text{Var}\,M[f]}{\delta f(\mathbf{P})\delta f(\mathbf{Q})}\right]_{f=f_{\text{opt}}} \varphi(\mathbf{P})\varphi(\mathbf{Q})\, d\mathbf{P}\, d\mathbf{Q} + \ldots \tag{18}$$

The term which is linear in $\varphi$ does not occur because $f_{\text{opt}}$ satisfies (7).

To evaluate the quadratic term in (18), it is sufficient to use functional derivatives and relations such as (8) and

$$\frac{\delta}{\delta f(\mathbf{P})} f(\mathbf{Q}) = \delta(\mathbf{P},\mathbf{Q}), \qquad \int \delta(\mathbf{P},\mathbf{Q})\varphi(\mathbf{P})\, d\mathbf{P} = \varphi(\mathbf{Q}). \tag{19}$$

A straightforward calculation yields our main technical result:

$$\text{Var}\,M[f_{\text{opt}} + \varphi]$$
$$= \frac{1}{\langle f_{\text{opt}}^2 \rangle} + \frac{1}{\langle f_{\text{opt}}^2 \rangle^3}\left\{\langle f_{\text{opt}}^2 \rangle \times \langle \overline{\varphi}^2 \rangle - \langle f_{\text{opt}} \times \overline{\varphi} \rangle^2\right\} + \ldots \tag{20}$$

where $\overline{\varphi} = \varphi - \langle \varphi \rangle$.

Non-negativity of the factor in curly braces follows from the standard Schwartz inequality.[d]

The first term on the r.h.s. of (20), $\langle f_{\text{opt}}^2 \rangle^{-1}$, is the absolute minimum for the variance of $M$ as established by the Rao-Cramer inequality [2], [3]. The latter is valid for all $\varphi$ and therefore is somewhat stronger than the result (20) which we have obtained only for sufficiently small $\varphi$. However, Eq. (20) gives a simple explicit description of the *deviation* from optimality and so makes possible the practical prescriptions presented below after Eq. (24).

It is convenient to talk about <u>*informativeness*</u> $I_f$ of a generalized moment $f$ with respect to the parameter $M$, defined by

$$I_f = \left(\text{Var}\,M[f]\right)^{-1}. \tag{21}$$

The informativeness of $f_{\text{opt}}$ is

$$I_{\text{opt}} = \langle f_{\text{opt}}^2 \rangle, \tag{22}$$

which corresponds to the Rao-Cramer limit. And the expansion (20) explicitly describes the deviations from the limit.

Informativeness is closely related to Fischer's information [2], [3] which, however, is an attribute of data whereas informativeness is a property of the moment.

---

[c] Rather surprisingly, none of a dozen or so textbooks and monographs on mathematical statistics that I checked (including a comprehensive practical guide [2] and a comprehensive mathematical treatment [3]) explicitly formulated the prescription in terms of the method of moments although equivalent formulas do occur e.g. in simple examples of specific estimates for the parameters of standard distributions; cf. [4].

[d] Note that the Schwartz inequality figures in standard rigorous proofs of the Rao-Cramer theorem.

THE METHOD OF QUASI-OPTIMAL MOMENTS. The fact that the solution (12) is the point of a quadratic minimum means that any moment $f_{\text{quasi}}$ which is close to (12) would be practically as good as the optimal solution (we will call such moments *quasi-optimal*). A quantitative measure of closeness is given by comparing the $O(1)$ and $O(\varphi^2)$ terms on the r.h.s. of (20):

$$\frac{\langle f_{\text{opt}}^2 \rangle \langle \overline{\varphi}^2 \rangle - \langle f_{\text{opt}} \overline{\varphi} \rangle^2}{\langle f_{\text{opt}}^2 \rangle^2} \ll 1, \tag{23}$$

where $\overline{\varphi} = f_{\text{quasi}} - \langle f_{\text{quasi}} \rangle - f_{\text{opt}}$.

The subtracted term in the numerator of (23) is non-negative, so dropping it results in a sufficient condition for Eq.(23). Furthermore, $\langle f_{\text{opt}} \overline{\varphi} \rangle$ would tend to be suppressed anyway whenever $f_{\text{quasi}}$ oscillates around $f_{\text{opt}}$. Assuming without loss of generality that $\langle f_{\text{quasi}} \rangle = 0$, we obtain the following convenient sufficient criterion:

$$\langle [f_{\text{quasi}} - f_{\text{opt}}]^2 \rangle \ll \langle f_{\text{opt}}^2 \rangle. \tag{24}$$

Taking into account this and Eq.(20) and denoting the usual $\sigma$ for $M$ for the optimal and quasi-optimal cases as $\sigma_{\text{opt}}$ and $\sigma_{\text{quasi}}$, respectively, one obtains:

$$\frac{\sigma_{\text{quasi}}}{\sigma_{\text{opt}}} \approx 1 + \frac{1}{2} \times \frac{\langle [f_{\text{quasi}} - f_{\text{opt}}]^2 \rangle}{\langle f_{\text{opt}}^2 \rangle}. \tag{25}$$

Now the method of quasi-optimal moments is as follows:

(i) construct a generalized moment $f_{\text{quasi}}$ using (12) as a guide so that $f_{\text{quasi}}$ were close to $f_{\text{opt}}$ in the integral sense of Eq.(24);

(ii) find $M$ by fitting $\langle f_{\text{quasi}} \rangle$ against $\langle f_{\text{quasi}} \rangle_{\text{exp}}$;

(iii) estimate the error for $M$ via (6);

(iv) $f_{\text{quasi}}$ may depend on $M$ to find which one can optionally use an iterative procedure starting from some value $M_0$ close to the true one.

For practical construction of quasi-optimal moments $f_{\text{quasi}}$ it is useful to reformulate (24) in terms of integrands. The explicit form for (24) is

$$\int d\mathbf{P}\, \pi(\mathbf{P}) [f_{\text{quasi}}(\mathbf{P}) - f_{\text{opt}}(\mathbf{P})]^2 \ll \int d\mathbf{P}\, \pi(\mathbf{P}) f_{\text{opt}}^2(\mathbf{P}). \tag{26}$$

As a rule of thumb, one would aim to minimize the bracketed expression on the l.h.s. of (26):

$$[f_{\text{quasi}}(\mathbf{P}) - f_{\text{opt}}(\mathbf{P})]^2 \ll f_{\text{opt}}^2(\mathbf{P}). \tag{27}$$

This should hold for "most" $\mathbf{P}$, i.e. taking into account the magnitude of $\pi(\mathbf{P})$: the inequality (27) may be relaxed in the regions which yield small contributions to the integral on the l.h.s. of (26).

THE EXAMPLE (14). Suppose the exact probability distribution differs from (14) by, say, a mild but complicated dependence of $\Gamma$ on $\mathbf{P}$ (as seen e.g. from some sort of perturbative calculations of theoretical corrections — a situation typical of high-energy physics problems [5]). Then the r.h.s. of (15) with a constant $\Gamma$ would correspond to a generalized moment which is only quasi-optimal but deviations from optimality may be practically negligible (depending on the "mildness" of the $\mathbf{P}$-dependence). So one could still use the moment given by the simplest formula (15) without significant loss of informativeness.

Alternatively, one could replace the analytical shape (15) by cruder piecewise constant or, better, piecewise linear approximations that would imitate the expression (15):

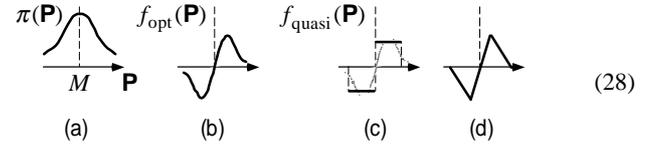

$$\tag{28}$$

In either case, the effect of non-optimality can be easily estimated via Eq.(25): the piecewise linear shape (d) deviates from optimality in the sense of (25) by a few per cent (in infinite domains, the slowly decreasing tails of the probability distribution may spoil this conclusion somewhat so one may wish to extend $f_{\text{quasi}}$ by additional linear pieces as well as insert flat linear pieces at the sharp peaks).

DISCUSSION. Eq.(27) allows one to talk about non-optimality of moments (i.e. their lower informativeness compared with $f_{\text{opt}}$) in terms of sources of non-optimality, i.e. the deviations of $f_{\text{quasi}}(\mathbf{P})$ from $f_{\text{opt}}(\mathbf{P})$ which give sizeable contributions to the l.h.s. of (24). The simplest example is when $f_{\text{opt}}$ is a continuous smoothly varying function whereas $f_{\text{quasi}}$ is a piecewise constant approximation (see (28), figure (c)). Then $f_{\text{quasi}}$ would usually deviate most from $f_{\text{opt}}$ near the discontinuities which, therefore, are naturally identified as sources of non-optimality. Then a natural way to improve $f_{\text{quasi}}$ is by "regulating" discontinuities via continuous (e.g. linear) interpolations.

Intuitively, one could think about sources of non-optimality as "leaks" through which information about $M$ is lost, and the improvement of $f_{\text{quasi}}$ would then correspond to patching up those leaks.

It is practically sufficient to take Eq.(12) at some value $M = M_0$ close to the true one (which is unknown anyway). This is usually possible in the case of precision measurements. One could also perform an iterative procedure for $M$ starting from $M_0$, then replacing $M_0$ with the value newly found, etc. — a procedure closely related to the optimization in the maximum likelihood method.

If $\pi(\mathbf{P})$ is given by a perturbation theory with increasingly complex but decreasingly important contributions, it is possible to use an approximate shape for the r.h.s. of (12) such as given by a few terms of a perturbative expansion in which the dependence on the parameter manifests itself. Theoretical updates of the complete $\pi(\mathbf{P})$ need not be always reflected in the quasi-optimal moments.

If the dimensionality of the space of events is not large then it may be possible to construct a suitable $f_{\text{quasi}}$ in a brute force fashion, i.e. build an interpolation formula for $\pi(\mathbf{P})$ for two or more values of $M$ near the value of interest, and perform the differentiation in $M$ numerically.

Also, one can use different expressions for $f_{\text{quasi}}$: e.g. perform a few first iterations with a simple shape for faster calculations and then switch to a more sophisticated interpolation formula for best precision.

SEVERAL PARAMETERS. With several parameters to be extracted from data there are the usual ambiguities due to reparametrizations but one can always define a moment per parameter according to (12). Then the informativeness (21) is a matrix (as is Fischer's information).

Since the covariance matrix of (quasi-)optimal moments is known (or can be computed from data), the mapping of the corresponding error ellipsoids for different confidence levels from the space of moments into the space of parameters is straightforward.

OPTIMAL MOMENTS AND THE LEAST SQUARES METHOD. The popular $\chi^2$ method makes a fit with a number of non-optimal moments (bins of a histogram). The histogramming implies a loss of information but the method is universal, verifies the probability distribution as a whole, and is implemented in standard software routines. On the other hand, the choice of $f_{quasi}$ requires a problem-specific effort but then the loss of information can in principle be made negligible by suitable adjustments of $f_{quasi}$.

The balance is, as usual, between the quality of custom solutions and the readiness of universal ones. However, once quasi-optimal moments are found, the quality of maximum likelihood method seems to become available at a lower computational cost.

The two methods are best regarded as complementary: One could first employ the $\chi^2$ method to verify the shape of the probability distribution and obtain the value of $M_0$ to be used as a starting point in the method of quasi-optimal moments in order to obtain the best final estimate for $M$.

An additional advantage of the method of quasi-optimal moments may be that some of the more sophisticated theoretical formalisms yield predictions for probability densities in the form of singular (and therefore not necessarily positive-definite everywhere) generalized functions (cf. the systematic gauge-invariant quantum-field-theoretic perturbation theory with unstable particles outlined in [6]). In such cases theoretical predictions for generalized moments (quasi-optimal or not) may exceed in quality predictions for probability densities, so that the use of the $\chi^2$ method would be somewhat disfavored compared with the method of quasi-optimal moments for the highest-precision measurements of unknown parameters.

Note that the data processing for the LEP1 experiments [5] has been performed in several iterations over several years and it would have been entirely possible to design, say, five quasi-optimal moments for the five parameters measured at the $Z$ resonance back in the '80s and to use them ever since.

CONCLUSIONS. It is clear that the method of quasi-optimal moments may be a useful addition to the data-processing arsenal e.g. in situations encountered in precision measurement problems in high-energy particle physics (cf. [5]) where one deals with $O(10^6)$ events and very complicated probability distributions obtained via quantum-field-theoretic perturbation theory so that the optimization involved in the maximum likelihood method is unfeasible. It also does not seem impossible to design universal software routines for a numerical construction of $f_{quasi}$ in the form of dynamically generated interpolation formulas.

Lastly, the usefulness of the concept of quasi-optimal moments is not limited to purely numerical situations: It also proved to be useful in a theoretical context of [1] as a guiding principle for studying an important class of data processing algorithms (the so-called jet finding algorithms).

ACKNOWLEDGMENTS. I thank Dima Bardin for a help with clarifying the bibliographic status of the concept of optimal moments. This work was supported in part by the Russian Foundation for Basic Research under grant 99-02-18365.